\documentclass[twocolumn,showpacs,preprintnumbers,amsmath,amssymb,prb]{revtex4}

\usepackage{graphicx}
\usepackage{dcolumn}
\usepackage{bm}

\begin{document}

\title{Spin qubits in antidot lattices}

\author{Jesper Pedersen,$^1$ Christian Flindt,$^1$ Niels Asger Mortensen,$^1$ and Antti-Pekka Jauho$^{1,2}$}
\affiliation{$^1$MIC -- Department of Micro and Nanotechnology,
             NanoDTU, Technical University of Denmark, Building 345east,
             DK-2800 Kongens Lyngby, Denmark\\
             $^2$Laboratory of Physics, Helsinki University of Technology, P.\ O.\ Box 1100, FI-02015 HUT, Finland}

\date{\today}
\begin{abstract}
We suggest and study designed defects in an otherwise periodic
potential modulation of a two-dimensional electron gas as an
alternative approach to electron spin based quantum information
processing in the solid-state using conventional gate-defined
quantum dots. We calculate the band structure and density of states
for a periodic potential modulation, referred to as an antidot
lattice, and find that localized states appear, when designed
defects are introduced in the lattice. Such defect states may form
the building blocks for quantum computing in a large antidot
lattice, allowing for coherent electron transport between distant
defect states in the lattice and tunnel coupling of neighboring
defect states with corresponding electrostatically controllable
exchange coupling between different electron spins.
\end{abstract}

\pacs{73.21.Cd, 75.30.Et, 73.22.-f}

\maketitle

\section{Introduction}

Localized electrons spins in a solid state structure have been
suggested as a possible implementation of a future device for
large-scale quantum information processing.\cite{Loss1998} Together
with single spin rotations, the exchange coupling between spins in
tunnel coupled electronic levels would provide a universal set of
quantum gate operations.\cite{Burkard1999} Recently, both of these
operations have been realized in experiments on electron spins in
double quantum dots, demonstrating electron spin resonance~(ESR)
driven single spin rotations\cite{Koppens2006} and electrostatic
control of the exchange coupling between two electron
spins.\cite{Petta:2005} Combined with the long coherence time of the
electron spin due to its weak coupling to the environment, and the
experimental ability to initialize a spin and reading it
out,\cite{Elzerman2004} four of DiVincenzo's five
criteria\cite{DiVincenzo2000a} for implementing a quantum computer
may essentially be considered fulfilled. This leaves only the
question of scalability experimentally unaddressed.

\begin{figure}[b!]
\begin{center}
\includegraphics[width=.95\linewidth]{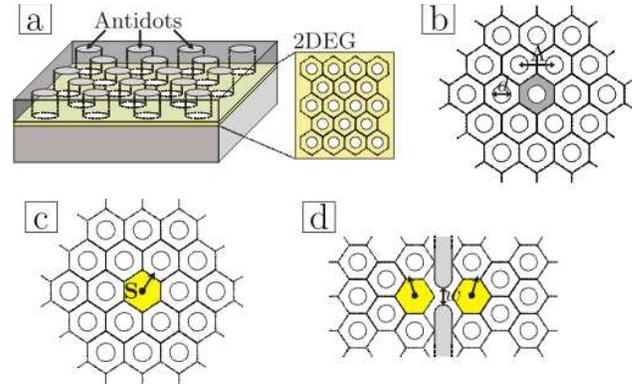}
\end{center}
\caption{(Color online) (a) Schematic illustration of a periodic
antidot lattice; antidots may, e.g., be fabricated using local
oxidation of a Ga[Al]As heterostructure. (b) Geometry of the
periodic antidot lattice with the Wigner--Seitz cell marked in gray
and the antidot diameter $d$ and lattice constant $\Lambda$
indicated. (c) A designed defect leads to the formation of defect
states in which an electron with spin $\mathbf{S}$ can reside. (d)
Tunnel coupled defects. The coupling can be controlled using a
split-gate with an effective opening denoted $w$.}
\label{fig:frontpage}
\end{figure}

\begin{figure*}
\begin{center}
\includegraphics[width=\textwidth]{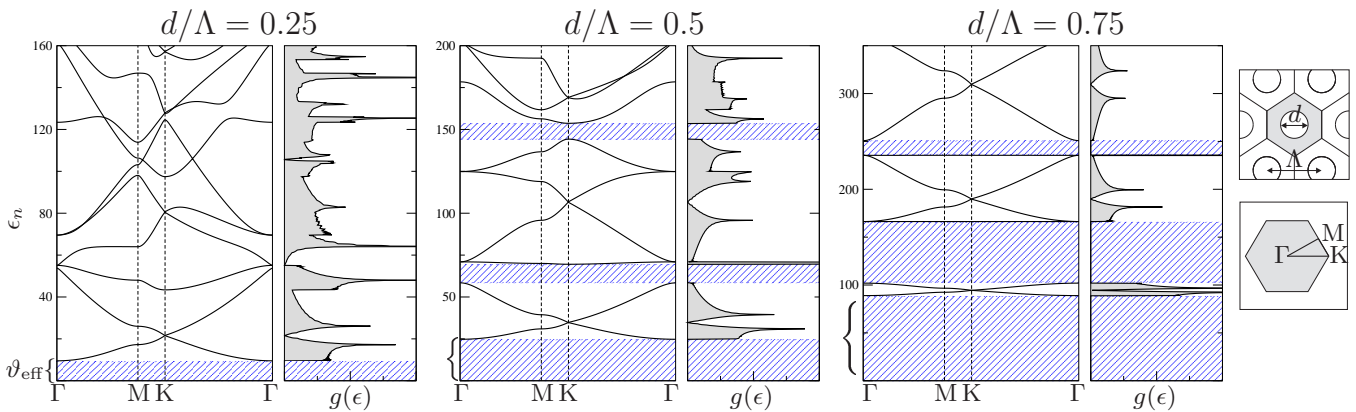}
\end{center}
\caption{(Color online) Band structures and densities of states
$g(\epsilon)$ of the periodic antidot lattice for three different
values of the relative antidot diameter $d/\Lambda$. Notice the
different energy scales for the three cases. On each band structure
the gap $\vartheta_\textrm{eff}$ is indicated, below which no states
exist for the periodic lattice. The band gaps and the gap below
$\vartheta_\textrm{eff}$ are highlighted as
hatched blue regions. Also shown is the periodic lattice structure with the
Wigner--Seitz cell indicated in gray, as well as the first Brillouin
zone (FBZ) with the three high-symmetry points and the irreducible FBZ indicated.}
\label{fig:periodic}
\end{figure*}

While large-scale quantum information processing with conventional
gate-defined quantum dots is a topic of ongoing theoretical
research,\cite{Taylor2005} we here suggest and study an alternative
approach based on so-called defect states that form at designed
defects in a periodic potential modulation of a two-dimensional
electron gas (2DEG) residing at the interface of a semiconductor
heterostructure.\cite{Flindt2005} One way of implementing the
potential modulation would be similar to the periodic antidot
lattices\cite{ENSSLIN1990,WEISS1993} that are now routinely
fabricated. Such lattices can be fabricated on top of a
semiconductor heterostructure using local oxidation techniques that
allow for a precise patterning of arrays of insulating islands, with
a spacing on the order of 100 nm, in the underlying
2DEG.\cite{Dorn2005} Even though the origin of these depletion spots
is not essential for our proposal, we refer to them as antidots, and
a missing antidot in the lattice as a defect. Alternative
fabrication methods include electron beam and photo lithography.
\cite{Luo:2006, Martin:2003}
In Ref.\ \onlinecite{Dorn2005} a square lattice consisting of
$20\times20=400$ antidots was patterned on an approximately 2.5
$\mu$m $\times$ 2.5 $\mu$m area, and the available fabrication
methods suggest that even larger antidot lattices with more than
1000 antidots and many defect states may be within experimental
reach.

The idea of using designed defects in antidot lattices as a possible
quantum computing architecture was originally proposed by some of us
in Ref.\ \onlinecite{Flindt2005}, where we presented simple
calculations of the single-particle level structure of an antidot
lattice with one or two designed defects. Here, we take these ideas
further and present detailed band structure and density of
states calculations for a periodic lattice, describe a resonant
tunneling phenomenon allowing for electron transport between distant
defects in the lattice, and calculate numerically the exchange
coupling between spins in two neighboring defects, showing that the suggested
architecture could be useful for spin-based quantum information
processing. The envisioned structure and the basic building blocks
are shown schematically in Fig.\ \ref{fig:frontpage}.

The paper is organized as follows: In Section
\ref{sec:periodiclattice} we introduce our model of the antidot
lattice and present numerical results for the band structure and
density of states of a periodic antidot lattice. In particular, we
show that the periodic potential modulation gives rise to band gaps
in the otherwise parabolic free electron band structure. In Section
\ref{sec:defectstates} we introduce a single missing antidot, a
defect, in the lattice and calculate numerically the eigenvalue
spectrum of the localized defect states that form at the location of
the defect. We develop a semi-analytic model that explains the
level-structure of the lowest-lying defect states. In Section
\ref{sec:tunnelcoupling} we consider two neighboring defect states
and calculate numerically the tunnel coupling between them. In
Section \ref{sec:resonantcoupling} we describe a principle for
coherent electron transport between distant defect states in the
antidot lattice, and illustrate this phenomenon by wavepacket
propagations. In Section \ref{sec:exchangecoupling} we present
numerically exact results for the exchange coupling between electron
spins in tunnel coupled defect states, before we finally in Section
\ref{sec:conclusions} present our conclusions.

\section{Periodic antidot lattice}
\label{sec:periodiclattice}

We first consider a triangular lattice of antidots with lattice
constant $\Lambda$ superimposed on a two-dimensional electron gas
(2DEG). The structure is shown schematically together with the
Wigner-Seitz cell in Fig.\ \ref{fig:frontpage}(b). While experiments
on antidot lattices are often performed in a semi-classical regime,
where the typical feature sizes and distances, e.g., the lattice
constant $\Lambda$, are much larger than the electron wavelength, we
here consider the opposite regime, where these length scales are
comparable, and a full quantum mechanical treatment is
necessary. In the effective-mass approximation we thus model the
periodic lattice with a two-dimensional single-electron Hamiltonian
reading
\begin{equation}
H = -\frac{\hbar^2}{2m^*}\nabla_\mathbf{r}^2+\sum_i V\left(\mathbf{r}-\mathbf{R}_i\right),\;
\mathbf{r}=(x,y),
\end{equation}
where $m^*$ is the effective mass of the electron and $V(\mathbf{r}-\mathbf{R}_i)$ is the
potential of the $i$'th antidot positioned at $\mathbf{R}_i$. We model each antidot as a
circular potential barrier of diameter $d$ so that
$V(\mathbf{r}-\mathbf{R}_i)=V_0$ for $| \mathbf{r}-\mathbf{R}_i | \leq d/2$ and zero otherwise.
In the limit $V_0\rightarrow \infty$ the eigenfunctions do not penetrate into the antidots,
and the Schr\"odinger equation may be written as
\begin{equation}
-\Lambda^2\nabla_{\mathbf{r}}^2\psi_n(\mathbf{r})=\epsilon_n\psi_n(\mathbf{r}),\label{eq:singlep}
\end{equation}
with the boundary condition $\psi_n=0$ in the antidots, and where we
have introduced the dimensionless eigenvalues
\begin{equation}
\epsilon_n=E_n\Lambda^22m^*/\hbar^2. \label{eq:conversionfactor}
\end{equation}
In the following we use parameter values typical of GaAs, for which
$\hbar^2/2m^* \simeq 0.6$~eV~nm$^2$ with $m^*=0.067m_e$, although
the choice of material is not essential. We have checked numerically
that our results are not critically sensitive to the approximation
$V_0\rightarrow\infty$, so long as the height is significantly
larger than any energies under consideration. All results presented
in this work have thus been calculated in this limit, for which the
simple form of the Schr\"odinger equation Eq.~(\ref{eq:singlep})
applies. In this limit, the band structures presented below are of a
purely geometrical origin. The band structure can be calculated by
imposing periodic boundary conditions and solving Eq.\
(\ref{eq:singlep}) on the finite domain of the Wigner--Seitz cell.
We solve this problem using a finite-element method.\cite{comsol}
The corresponding density of states is calculated using the linear
tetrahedron method in its symmetry corrected
form.\cite{Lehmann1972,HAMA1990,Pedersen2007a}

In Fig.\ \ref{fig:periodic} we show the band structure and density
of states of the periodic antidot lattice for three different values
of the relative antidot diameter $d/\Lambda$. We note that an
increasing antidot diameter raises the kinetic energy of the Bloch
states due to the increased confinement and that several band gaps
open up. We have indicated the gap $\vartheta_\mathrm{eff}$ below
which no states exist for the periodic structure. We shall denote as
\emph{band gaps} only those gaps occurring between two bands, and
thus we do not refer to the gap below $\vartheta_\mathrm{eff}$ as a
band gap in the following. This is motivated by the difference in
the underlying mechanisms responsible for the gaps: While the band
gaps rely on the periodicity of the antidot lattice, similar to
Bragg reflection in the solid state, the gap below
$\vartheta_\mathrm{eff}$ represents an averaging of the potential
landscape generated by the antidots, and is thus robust against
lattice disorder as we have also checked
numerically.\cite{Pedersen:2007} The lowest band gap is thus present
for $d/\Lambda > 0.35$ while the higher-energy band gap only
develops for $d/\Lambda > 0.45$. As the antidot diameter is
increased, several flat bands appear with
$\nabla_\mathbf{k}\epsilon_n(\mathbf{k})\simeq 0$, giving rise to
van Hove singularities in the corresponding density of states.

\section{Defect states}
\label{sec:defectstates}

We now introduce a defect in the lattice by leaving out a single
antidot. Topologically, this structure resembles a planar 2D
photonic crystal, and relying on this analogy we expect one or more
localized defect states to form inside the
defect.\cite{Mortensen2005} The gap $\vartheta_\textrm{eff}$ indicated 
in Fig.\ \ref{fig:periodic} may be considered as the
height of an effective two-dimensional potential surrounding the
defect, and thus gives an upper limit to the existence of defect
states in this gap. Similar states are expected to form in the
band gaps of the periodic structure, which are highlighted in Fig.\
\ref{fig:periodic}. As defect states decay to zero far from the
location of the defect, we have a large freedom in the way we
spatially truncate the problem at large distances. For simplicity we
use a super-cell approximation, but with $\psi=0$ imposed on the
edge, thus leaving Eq.\ (\ref{eq:singlep}) a Hermitian eigenvalue
problem which we may conveniently solve with a finite-element
method.\cite{comsol} Other choices, such as periodic boundary
conditions, do not influence our numerical results. The size of the
super-cell has been chosen sufficiently large, such that the results
are unaffected by a further increase in size.

\begin{figure}
\begin{center}
\includegraphics[width=\linewidth, trim=0 0 -25 0, clip]{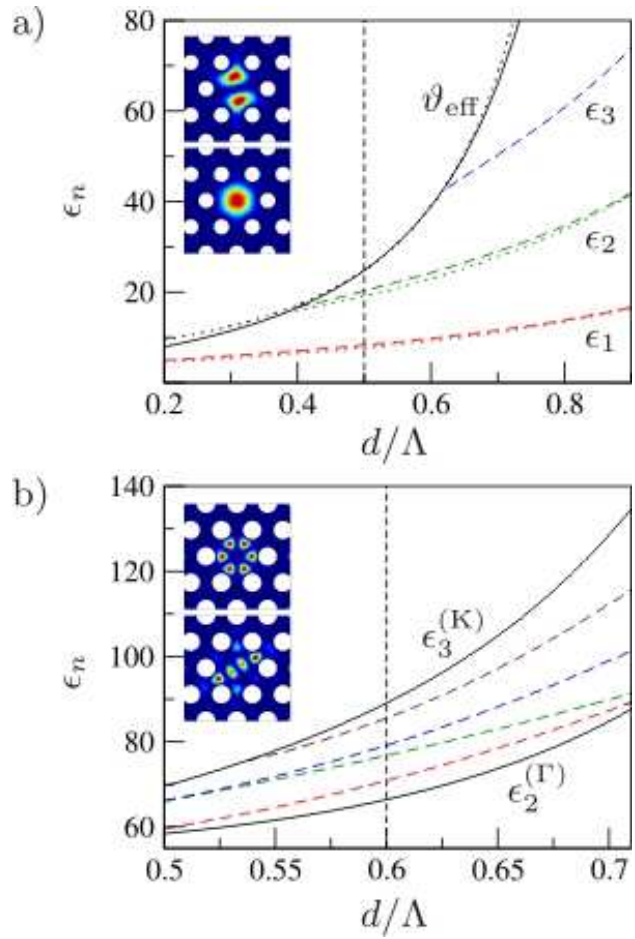}
\end{center}
\caption{(Color online) Energy spectrum for a single defect. The
(dimensionless) eigenvalues corresponding to localized states are
shown as a function of the relative antidot diameter $d/\Lambda$.
For a given choice of $\Lambda$, the eigenvalues can be converted to
meV using Eq.\ (\ref{eq:conversionfactor}). (a) Energy spectrum for
defect states residing in the gap below $\vartheta_\textrm{eff}$.
The full line indicates the height $\vartheta_\textrm{eff}$ of the
effective potential in which the localized states reside. The dotted
lines are the approximate expressions given by
Eqs.~(\ref{eq:veff}),~(\ref{eq:eps1}) and (\ref{eq:eps2}). The
approximate results for $\epsilon_1$ are in almost perfect agreement
with the numerical calculations. (b) Energy spectrum for the defect
states residing in the lowest band gap region. The full lines
indicate the band gap edges of the periodic structure,
$\epsilon_3^{(\textrm{K})}$ and $\epsilon_2^{(\Gamma)}$, giving
upper and lower limits to the existence of bound states. The inset
in both figures show the localized states corresponding to the two
lowest energy eigenvalues indicated by the dashed vertical lines.
The absolute square is shown.} \label{fig:defectstates}
\end{figure}

In the insets of Fig.\ \ref{fig:defectstates}(a) we show the
calculated eigenfunctions corresponding to the two lowest energy
eigenvalues for a relative antidot diameter $d/\Lambda=0.5$. As
expected, we find that defect states form that to a high degree are
localized within the defect. The second-lowest eigenvalue is
two-fold degenerate and we only show one of the corresponding
eigenstates. The figure shows the energy eigenvalues of the defect
states as a function of the relative antidot diameter together with
the gap $\vartheta_\textrm{eff}$. As this effective potential is
increased, additional defect states become available and we may thus
tune the number of levels in the defect by adjusting the relative
antidot diameter. In particular, we note that for $d/\Lambda\lesssim
0.42$ only a single defect state forms. As the sizes of the antidots
are increased, the confinement of the defect states becomes stronger,
leading to an increase in their energy eigenvalues. For GaAs with
$d/\Lambda=0.5$ and $\Lambda=75$~nm the energy splitting of the two
lowest defect states is approximately 1.1~meV, which is much larger
than $k_BT$ at subkelvin temperatures, and the level structure is
thus robust against thermal dephasing.

In Fig.~\ref{fig:defectstates}(b) we show similar results for defect
states residing in the lowest band gap of the periodic structure.
While the states residing below $\vartheta_\textrm{eff}$ resemble
those occurring due to the confining potential in conventional
gate-defined quantum dots, these higher-lying states are of a very
different nature, being dependent on the periodicity of the
surrounding lattice. For the band gaps, the existence of bound
states is limited by the relevant band edges as indicated in the
figure. As the size of the band gap is increased, additional defect
states become available and we may thus also tune the number of
levels residing in the band gaps by adjusting the relative antidot
diameter.

Because the formation of localized
states residing below $\vartheta_\textrm{eff}$ depends only on the
existence of the effective potential surrounding the defect, the
formation of such states is not critically dependent on perfect
periodicity of the surrounding lattice, which we have checked
numerically.\cite{Pedersen:2007} Also, the lifetimes of the states
due to the finite size of the antidot lattice are of the order of
seconds even for a relatively small number of rings of antidots
surrounding the defect.\cite{Flindt2005} However, the localized
states residing in the band gaps are more sensitive to lattice
disorder, since they rely more crucially on the periodicity of the
surrounding lattice. Introducing disorder may induce a finite
density of states in the band gaps of the periodic structure and thus
significantly decrease the lifetimes of the localized states
residing in this region.

In order to gain a better understanding of the level-structure of
the defect states confined by $\vartheta_\textrm{eff}$ we develop a
semi-analytic model for $\vartheta_\textrm{eff}$ and the
corresponding defect states. We first note that the effective
potential $\vartheta_\textrm{eff}$ is given by the energy of the
lowest Bloch state at the $\Gamma$ point of the periodic lattice. At
this point $\mathbf{k}=\mathbf{0}$ and Bloch's theorem reduces to an
ordinary Neumann boundary condition on the edge of the Wigner--Seitz
cell. This problem may be solved using a conformal mapping, and we
obtain the expression\cite{Mortensen2006}
\begin{equation}
\vartheta_\textrm{eff} \simeq \left(\mathcal{C}_1 + \frac{\mathcal{C}_2}{\mathcal{C}_3-d/\Lambda} \right)^2, \label{eq:veff}
\end{equation}
where $\mathcal{C}_1\simeq -0.2326$, $\mathcal{C}_2\simeq 2.7040$
and $\mathcal{C}_3\simeq 1.0181$ are given by expressions involving
the Bessel functions $Y_0$ and $Y_1$. \cite{Mortensen2006} We now
consider the limit of $d/\Lambda \rightarrow 1$ and note that in
this case the defect states residing below $\vartheta_\mathrm{eff}$
are subject to a potential which we may approximate as an infinite
two-dimensional spherical potential well with radius $\Lambda-d/2$.
The lowest eigenvalue for this problem is
$\epsilon_1^{(\infty)}=\Lambda^2\alpha_{0,1}^2/(\Lambda-d/2)^2$,
where $\alpha_{0,1}\simeq 2.405$ is the first zero of the zeroth
order Bessel function. This expression yields the correct scaling
with $d/\Lambda$, but is only accurate in the limit of $d/\Lambda
\rightarrow 1$. We correct for this by considering the limit of
$d/\Lambda\rightarrow 0$, in which we may solve the problem using
ideas developed by Glazman \emph{et al.} in studies of quantum
conductance through narrow constrictions. \cite{Glazman1988} The
problem may be approximated as a two-dimensional spherical potential
well of height $\pi^2$ and radius $\Lambda$. The lowest eigenvalues
$\epsilon_1^{(\pi^2)}$ of this problem is the first root of the
equation
\begin{equation}
\sqrt{\epsilon_1^{(\pi^2)}}\frac{J_1\left(\sqrt{\epsilon_1^{(\pi^2)}}\right)}{J_0\left(\sqrt{\epsilon_1^{(\pi^2)}}\right)}
=
\sqrt{\pi^2-\epsilon_1^{(\pi^2)}}\frac{K_1\left(\sqrt{\pi^2-\epsilon_1^{(\pi^2)}}\right)}{K_0\left(\sqrt{\pi^2-\epsilon_1^{(\pi^2)}}\right)},
\label{eq:root}
\end{equation}
where $J_i$($K_i$) is the $i$'th order Bessel function of the first
(second) kind. If the height of the potential well $\pi^2$ is much
larger than the energy eigenvalues, the first root would simply be
$\alpha^2_{0,1}$. Lowering the confinement must obviously shift down
the eigenvalue, and in the present case we find that
$\epsilon_1^{(\pi^2)}\simeq \pi$. By expanding the equation to first
order in $\sqrt{\epsilon_1^{(\pi^2)}}$ around $\sqrt{\pi}$ we may
solve the equation to obtain $\epsilon_1^{(\pi^2)}\simeq 3.221$,
which is in excellent agreement with a full numerical solution of
Eq.~(\ref{eq:root}). Correcting for the low-$d/\Lambda$ behavior we
thus find the approximate expression for the lowest energy
eigenvalue\cite{Flindt2005}
\begin{eqnarray}
\epsilon_1 &\simeq& \epsilon_1^{(\infty)} - \lim_{d/\Lambda\rightarrow 0}\epsilon_1^{(\infty)}+\epsilon_1^{(\pi^2)} \nonumber \\
&=& \epsilon_1^{(\pi^2)} + \frac{\left(4-d/\Lambda\right)d/\Lambda}{\left(2-d/\Lambda\right)^2}\alpha_{0,1}^2.\label{eq:eps1}
\end{eqnarray}
A similar analysis leads to an approximate expression for the first
excited state $\epsilon_2$. This mode has a finite angular momentum
of $\pm 1$ and a radial $J_1$ solution yields
\begin{eqnarray}
\epsilon_2 \simeq \epsilon_2^{(\pi^2)} + \frac{\left(4-d/\Lambda\right)d/\Lambda}{\left(2-d/\Lambda\right)^2}\alpha_{1,1}^2,\label{eq:eps2}
\end{eqnarray}
where $\epsilon_2^{(\pi^2)}\simeq 7.673$ is the second-lowest
eigenvalue of the two-dimensional spherical potential well of height
$\pi^2$ and radius $\Lambda$, which can be found from an equation
very similar to Eq.~(\ref{eq:root}). The first root of the
first-order Bessel function is $\alpha_{1,1}\simeq 3.832$. The
scaling of the two lowest eigenvalues with $d/\Lambda$ is thus
approximately the same. The approximate expressions are indicated by
the dotted lines in Fig.~\ref{fig:defectstates}, and we note an
excellent agreement with the numerical results. We remark that the
filling of the defect states can be controlled using a metallic back
gate that changes the electron density and thus the occupation of
the different defect states.\cite{footnote:filling}

\section{Tunnel coupled defect states}
\label{sec:tunnelcoupling}

Two closely situated defect states can have a finite tunnel
coupling, leading to the formation of hybridized defect states. The
coupling between the two defects may be tuned via a metallic split
gate defined on top of the 2DEG in order to control the opening
between the two defects. As the voltage is increased the opening is
squeezed, leading to a reduced overlap between the defect states. We
model such a split gate as an infinite potential barrier shaped as
shown in Fig.~\ref{fig:frontpage}(d). Changing the applied voltage
effectively leads to a change in the relative width $w/\Lambda$ of
the opening, which we take as a control parameter in the following.
If we consider just a single level in each defect we can calculate
the tunnel matrix element as $|\tau|=(\epsilon_+-\epsilon_-)/2$
where $\epsilon_\pm$ are the eigenenergies of the bonding and
anti-bonding states, respectively, of the double defect. In the
following, we calculate the tunnel coupling between two defect
states lying below $\vartheta_\textrm{eff}$, but the analysis
applies equally well to defect states lying in the band gaps.

\begin{figure}
\begin{center}
\includegraphics[width=\linewidth, trim=0 0 -25 0, clip]{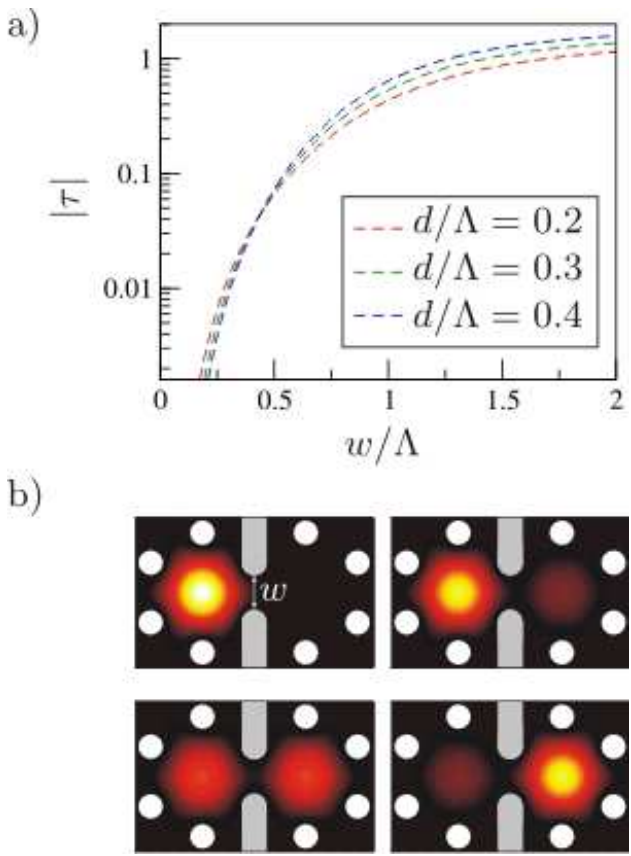}
\end{center}
\caption{(Color online) (a) The (dimensionless) tunnel coupling
$|\tau|$ as a function of the relative split gate constriction width
$w/\Lambda$ for three different values of $d/\Lambda$ in the
single-level regime. For a given choice of $\Lambda$, the tunnel
couplings can be converted to meV using Eq.\
(\ref{eq:conversionfactor}). (b) Time propagation of an electron
initially prepared in the left defect state for $d/\Lambda=0.4$ and
$w/\Lambda=0.6$. The absolute square of the initial wavefunction is
shown in the upper left panel. The following panels show the state
after a time span of $T/8$, $2T/8$ and $3T/8$, respectively, where
$T$ is the oscillation period.} \label{fig:tunnel}
\end{figure}

In Fig.~\ref{fig:tunnel} we show the tunnel matrix element $|\tau|$
as a function of the relative gate constriction width $w/\Lambda$
for three different values of $d/\Lambda$ in the single-level regime
of each defect, i.e., $d/\Lambda\lesssim 0.42$. As expected, the
tunnel coupling grows with increasing constriction width due to the
increased overlap between the defect states. A saturation point is
reached when the constriction width is on the order of the diameter
of the defect states, after which the overlap is no longer increased
significantly. An electron prepared in one of the defect states will
oscillate coherently between the two defect states with a period
given as $T=\pi\hbar/|\tau|$, which for GaAs with $\Lambda=75$~nm,
$d/\Lambda=0.4$ and $w/\Lambda=0.6$ implies an oscillation time of
$T\simeq 0.14$~ns. A numerical wavepacket propagation of an electron
initially prepared in the left defect state is shown in
Fig.~\ref{fig:tunnel}(b), confirming the expected oscillatory
behavior. With a finite tunnel coupling between two defect states,
two electron spins trapped in the defects will interact due to the
exchange coupling, to which we return in Section
\ref{sec:exchangecoupling}.

\section{Resonant coupling of distant defect states}
\label{sec:resonantcoupling}

\begin{figure}
\begin{center}
\includegraphics[width=\linewidth]{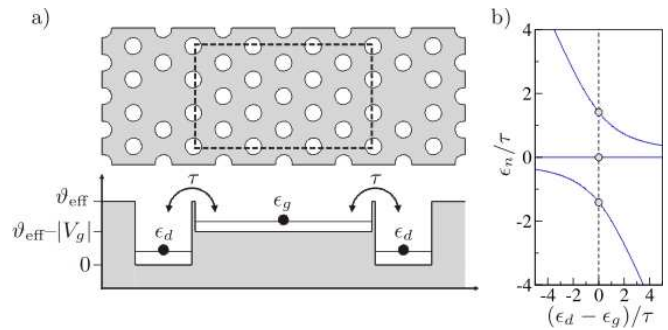}
\end{center}
\caption{(a) The structure considered for resonant coupling of distant defect
states; two defects separated by a central line of $N=3$ antidots, with a
central back gate $V_g$ controlling the potential square well in the region
marked with dashed lines.
A simple three-level model of the system is
illustrated below. (b) The eigenvalue spectrum of the three-level model.
The dashed line marks the point of resonance.}
\label{fig:geomRes}
\end{figure}

With a large antidot lattice and several defect states it may be
convenient with quantum channels along which coherent electron
transport can take place, connecting distant defect states. In Refs.\
\onlinecite{Nikolopoulos2004} and \onlinecite{Nikolopoulos2004a} it
was suggested to use arrays of tunnel coupled quantum dots as a
means to obtain high-fidelity electron transfer between two distant
quantum dots. We have applied this idea to an array of tunnel
coupled defect states and confirmed that this mechanism may be used for coherent
electron transport between distant defects in an antidot
lattice.\cite{Pedersen:2007} This approach, however, relies on
precise tunings of the tunnel couplings between each defect in the
array, which may be difficult to implement experimentally. Instead,
we suggest an alternative approach based on a resonant coupling
phenomenon inspired by similar ideas used to couple light between
different fiber cores in a photonic crystal
fiber.\cite{Skorobogatiy2006a,Skorobogatiy2006}

We consider two defects separated by a central line of $N$ antidots
and a central back gate $V_g$ in the region between the defects, as
shown in Fig.~\ref{fig:geomRes}. Again, we consider defect states
residing below $\vartheta_\textrm{eff}$, but the principle described
here may equally well be applied to defect states in the band gaps.
Using the back gate, the potential between
the two defects can be controlled locally. If the potential is
lowered below $\vartheta_\textrm{eff}$, a discrete spectrum of
standing-wave solutions forms between the two defects. In the
following we denote the energy of one of these standing-wave
solutions by $\epsilon_g$, while the energy of the two defect states
is assumed to be identical and is denoted $\epsilon_d$. A simple
three-level analysis of this system, as illustrated in
Fig.~\ref{fig:geomRes}, reveals that by tuning the back gate so that
the levels are aligned, $\epsilon_g=\epsilon_d$, a resonant coupling
between the two distant defects occurs, characterized by a symmetric
splitting of the three lowest eigenvalues into
$\epsilon_0=\epsilon_d$ and
$\epsilon_\pm=\epsilon_d\pm\sqrt{2}|\tau|$, where $|\tau|$ is the
tunnel coupling between the defects and the standing-wave solution
in the central back gate region. If an electron is prepared in one
of the defects states, it will oscillate coherently between the two
defects with an oscillation period of $T=\sqrt{2}\pi\hbar/|\tau|$.
By turning off the back gate at time $t=T/2$ we may thereby trap the
electron in the opposite defect which may by situated a distance an
order of magnitude larger than the lattice constant away from the
other defect.

\begin{figure}
\begin{center}
\includegraphics[width=\linewidth, trim=0 0 -25 0, clip]{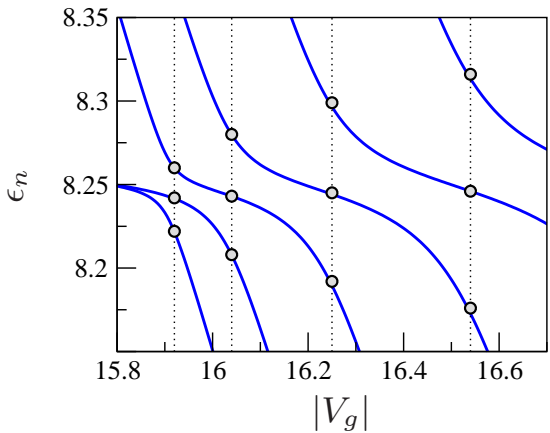}
\end{center}
\caption{(Color online) Energy eigenvalues as a function of the magnitude $|V_g|$
of the back gate for the structure illustrated in Fig.~\ref{fig:geomRes}
for $d/\Lambda=0.5$ and a central line of $N=7$ antidots separating the two
defects. The resonances are marked with dotted lines and characterized by a
symmetric splitting of the eigenvalues. }
\label{fig:resN7}
\end{figure}

\begin{figure}
\begin{center}
\includegraphics[width=1\linewidth, trim=0 0 0 0, clip]{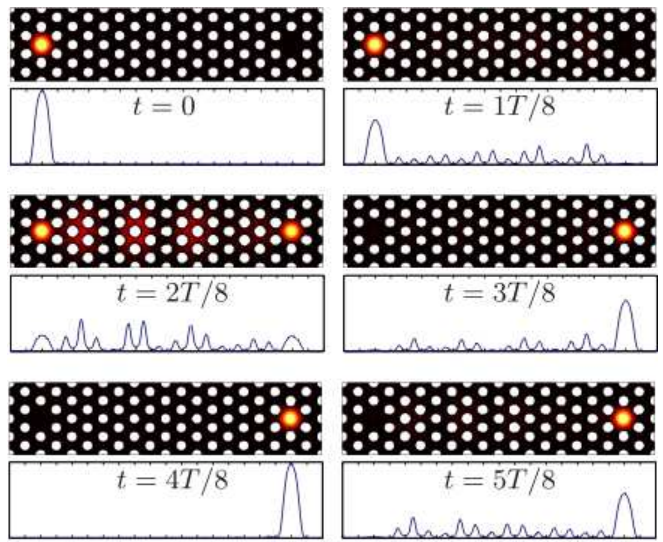}
\end{center}
\caption{(Color online) Numerical time propagation of
an electron initially prepared in the left defect of the structure illustrated
in Fig.~\ref{fig:geomRes}(a) and corresponding to the results of
Fig.~\ref{fig:resN7} with $|V_\mathrm{g}|\simeq 16.54$. The charge densities
$\rho(x,y)$ are shown in the upper panels, while the lower panels show
$\int dy\rho(x,y)$. The oscillation period is denoted $T$.}
\label{fig:resProp}
\end{figure}

In Fig.~\ref{fig:resN7} we show the numerically calculated
eigenvalues as a function of the depth $|V_g|$ of the central
potential square well of the structure illustrated in
Fig.~\ref{fig:geomRes} for $d/\Lambda=0.5$ and a central line of
$N=7$ antidots separating the two defects. Contrary to the simple
three-level model, several resonances now occur as the back gate is
lowered, corresponding to coupling to different standing-wave
solutions in the multi-leveled central region. The energy splitting
at resonance is larger when the defect states couple to higher-lying
central states due to a large overlap between the defect states and
the central standing-wave solution. In Fig.~\ref{fig:resProp} we
show a numerical time propagation of an electron initially prepared
in the left defect, confirming the oscillatory behavior expected
from the simple model. For GaAs and $\Lambda=75$~nm the results
indicate an oscillation period of $T\simeq 0.16$~ns for the time
propagation illustrated. The resonant phenomenon relies solely on
the level alignment $\epsilon_g=\epsilon_d$ and on the symmetry
condition that both defect states have the same energy and magnitude
of tunnel coupling to the standing wave solution in the central
region. It is in principle independent of the number of antidots $N$
separating the two defects, but in practice this range is limited by
the coherence length of the sample and the fact that the levels of
the central region grow too dense if $N$ becomes
large.\cite{footnote:rctime}
We have checked numerically
that resonant coupling of defect levels below
$\vartheta_\textrm{eff}$ is robust against lattice
disorder.\cite{Pedersen:2007}

\section{Exchange coupling}
\label{sec:exchangecoupling}

So far we have only considered the single-particle electronic
level-structure of the antidot lattice. However, as mentioned in the
introduction, the exchange coupling between electron spins is a
crucial building block for a spin based quantum computing
architecture, and in fact suffices to implement a universal set of
quantum gates. \cite{DiVincenzo2000} The exchange coupling is a
result of the Pauli principle for identical fermions, which couples
the symmetries of the orbital and spin degrees of freedom. If the
orbital wavefunction of the two electrons is symmetric (i.e.\
preserves sign under particle-exchange), the spins must be in the
antisymmetric singlet state, while an antisymmetric orbital
wavefunction means that the spins are in a symmetric triplet state.
One may thereby map the splitting between the ground state energy
$E_S$ of the symmetric orbital subspace and the ground state energy
$E_A$ of the antisymmetric orbital subspace onto an effective
Heisenberg spin Hamiltonian
$\mathcal{H}=J\mathbf{S}_1\!\cdot\!\mathbf{S}_2$, where $J=E_A-E_S$
is the exchange coupling between the two spins, $\mathbf{S}_1$ and
$\mathbf{S}_2$. The implementation of quantum gates based on the
exchange coupling requires that $J$ can be varied over several
orders of magnitude in order to effectively turn the coupling on and
off. In this section we present numerically exact results for the
exchange coupling between two electron spins residing in tunnel
coupled defects as those illustrated in Fig.~\ref{fig:frontpage}(d).

The Hamiltonian  of two electrons in two tunnel coupled defects may
be written as
\begin{equation}
H(\mathbf{r}_1, \mathbf{r}_2) =
h(\mathbf{r}_1)+h(\mathbf{r}_2)+C(\mathbf{r}_1,\mathbf{r}_2),
\end{equation}
where
\begin{equation}
C(\mathbf{r}_1,\mathbf{r}_2)=\frac{e^2}{4\pi\epsilon_r\epsilon_0}\frac{1}{|\mathbf{r}_1-\mathbf{r}_2|}
\end{equation}
is the Coulomb interaction and the single-electron Hamiltonians are
\begin{equation}
h(\mathbf{r}_i)=\frac{(\mathbf{p}_i+e\mathbf{A})^2}{2m^*}+V(\mathbf{r}_i)+\frac{1}{2}g\mu_BB S_{z,i}\, ,\,
i=1,2,\label{eq:hi}
\end{equation}
where $V(\mathbf{r})$ is the potential due to the antidots and the
coupled defects. As previously, we model the antidots and the split
gate as potential barriers of infinite height, and use
finite-element methods to solve the single-electron problem defined by
Eq.~(\ref{eq:hi}). A
Zeeman field $B\hat{\mathbf{z}}$ applied perpendicularly to the
electron gas splits the spin states, and we choose a corresponding
vector potential reading
$\mathbf{A}=B(-y\hat{\mathbf{x}}+x\hat{\mathbf{y}})/2$.

\begin{figure}
\begin{center}
\includegraphics[width=\linewidth, trim=0 0 -25 0, clip]{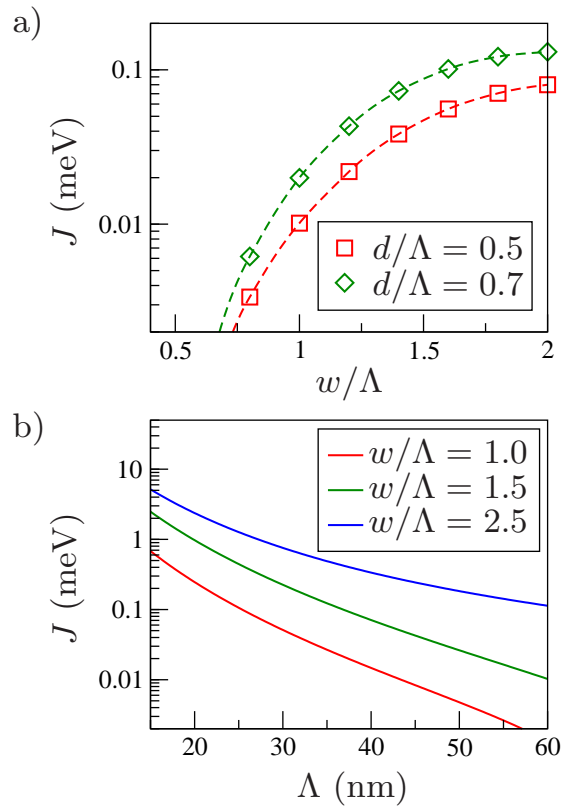}
\end{center}
\caption{(Color online) Exchange coupling $J$ for a double defect structure.
(a) Exchange coupling as a function of the relative split gate constriction
width $w/\Lambda$ for two different values of the relative antidot diameter and
a lattice constant $\Lambda=45$~nm.
(b) Exchange coupling as a function of the lattice constant $\Lambda$ for
three different values of the relative split gate constriction width.}
\label{fig:Jvsw}
\end{figure}
In order to calculate the exchange coupling $J$ we employ a recently
developed method for numerically exact finite-element calculations
of the exchange coupling:\cite{Pedersen2007} The full two-electron
problem is solved by expressing the two-electron Hamiltonian in a
basis of product states of single-electron solutions obtained using
a finite element method.\cite{comsol} The Coulomb matrix elements
are evaluated by expanding the single-electron states in a basis of
2D Gaussians,\cite{Helle:2005} and the two-particle Hamiltonian
matrix resulting from this procedure may then be diagonalized in the
subspaces spanned by the symmetric and antisymmetric product states,
respectively, to yield the exchange coupling. The details of the
numerical method are described
elsewhere.\cite{Pedersen2007,Pedersen:2007} The results presented
below have all been obtained with a sufficient size of the 2D
Gaussian basis set as well as the number of single-electron
eigenstates, such that a further increase does not change the
results.\cite{footnote:expansion}

\begin{figure}
\begin{center}
\includegraphics[width=\linewidth, trim=0 0 -25 0, clip]{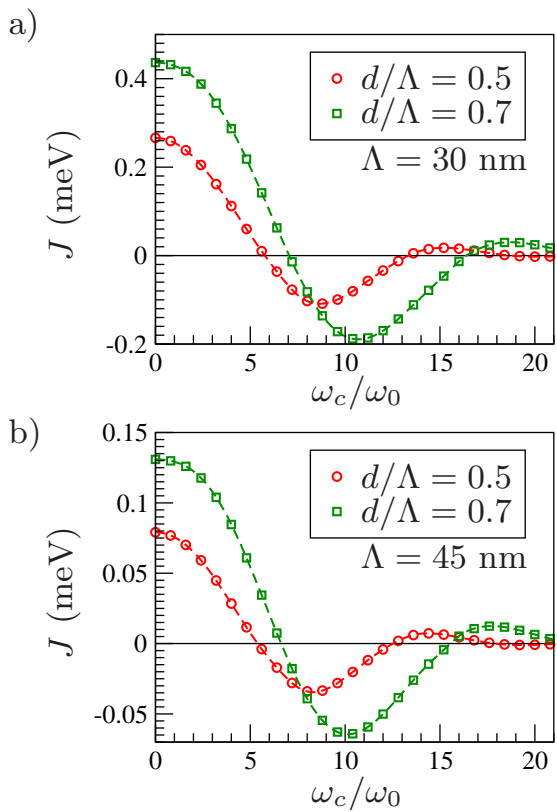}
\end{center}
\caption{(Color online) Exchange coupling $J$ for a double defect structure
as a function of $\omega_c/\omega_0$, where $\omega_c=eB/m^*c$ and
$\omega_0=\hbar/(2m^*\Lambda^2)$. Results are shown
for a relative split gate constriction width $w/\Lambda=2$,
and two different values of the relative antidot diameter $d/\Lambda$.
The lattice constant is (a) $\Lambda=30$~nm and (b) $\Lambda=45$~nm.}
\label{fig:JvsB}
\end{figure}

In Fig.~\ref{fig:Jvsw} we show the calculated exchange coupling for
a double defect structure. The exchange coupling varies by several
orders of magnitude as the split gate constriction width is
increased, showing that electrostatic control of the exchange
coupling in an antidot lattice is possible, similarly to the
principles proposed\cite{Burkard1999} and experimentally
realized\cite{Petta:2005} for double quantum dots. Just as the
tunnel coupling, the exchange coupling reaches a saturation point
when the split gate constriction width is on the order of the
diameter of the defect states. This is to be expected since the
exchange coupling in the Hubbard approximation is proportional to
the square of the tunnel coupling.\cite{Burkard1999} As illustrated
in Fig.~\ref{fig:Jvsw}(b), the exchange coupling is highly dependent
on the lattice constant, increasing several orders of magnitude as
the lattice constant is decreased from $60$~nm to $20$~nm. This is
in part due to the overall increase in the energies of the eigenstates 
and the between them with increased confinement, but also due to a
decrease in the ratio of the Coulomb interaction strength to the
confinement strength. As the relative strength of the Coulomb
interaction is decreased, the defect states are effectively moved
closer together, resulting in an increase in the exchange coupling.

The exchange coupling is also highly dependent on magnetic fields
applied perpendicularly to the plane of the
electrons.\cite{Burkard1999} In Fig.~\ref{fig:JvsB} we show the
exchange coupling as a function of $\omega_c/\omega_0$ where
$\omega_c=eB/m^*c$ and we define
$\omega_0=\frac{\hbar}{2m^*\Lambda^2}$. For GaAs
$\omega_c/\omega_0\simeq 0.00104$~T$^{-1}$nm$^{-2}\cdot\Lambda^2B$.
As expected, the results of Fig.~\ref{fig:JvsB} are very similar to
those obtained for double quantum dots.
\cite{Burkard1999,Helle:2005} In all cases we note an initial
transition from the anti-ferromagnetic ($J>0$) to the ferromagnetic
($J<0$) regime of exchange coupling, followed by a return to
positive values of the exchange coupling at higher magnetic fields.
The initial transition to negative exchange coupling is caused by
long-range Coulomb interactions.\cite{Burkard1999} As the magnetic
field is increased further, magnetic confinement becomes dominant,
compressing the orbits and thus reducing the overlap between the
single-defect wave functions. This leads to a strong reduction of
the magnitude of the exchange coupling. Due to the increased
confinement strength for smaller lattice constants $\Lambda$, these
transitions occur at larger magnetic fields. The same is the case
for the larger relative antidot diameters, in which the ratio of
magnetic confinement to confinement due to the antidots is reduced.
We have only considered the case of a large constriction width
$w/\Lambda=2$, since this regime of relatively large exchange
coupling is the most interesting for practical purposes. For small
values of $w/\Lambda$ we expect to find results similar to those
obtained in the limit of large interdot distances for double quantum
dot systems.\cite{Burkard1999}

\section{Conclusions}
\label{sec:conclusions}

In conclusion, we have suggested and studied an alternative
candidate for spin based quantum information processing in the
solid-state, namely defect states forming at the location of
designed defects in an otherwise periodic potential modulation of a
two-dimensional electron gas, here referred to as an antidot
lattice. We have performed numerical band structure and density of
states calculations of a periodic antidot lattice, and shown how
localized defect states form at the location of designed defects.
The antidot lattice allows for resonant coupling of distant defect
states, enabling coherent transport of electrons between distant
defects. Finally, we have shown that electrostatic control of the
exchange coupling between electron spins in tunnel coupled defect
states is possible, which is an essential ingredient for spin based
quantum computing. Altogether, we believe that designed defects in
antidot lattices provide several prerequisites for a large quantum
information processing device in the solid state.

\begin{acknowledgments}
We thank A.\ Harju for helpful advice during the development of our
numerical routines, and T.\ G.\ Pedersen for fruitful discussions
during the preparation of this manuscript. APJ is grateful to the
FiDiPro program of the Finnish Academy for support during the final
stages of this work.
\end{acknowledgments}


\begin{thebibliography}{26}
\bibitem[{foot()}]{footnote:filling}
\bibinfo{note}{In practice, reliable
single-electron filling may pose a serious experimental challenge
which could require further optimization of the architecture
presented here.}
\bibitem[{foot()}]{footnote:rctime}
\bibinfo{note}{Another experimental challenge relates to the $RC$
switching time of the back gate which grows with length, making it
harder to control on short time scales.}
\bibitem[{foot()}]{footnote:expansion}
\bibinfo{note}{Because we use an expansion in localized
single-particle product states our method is most reliable for
systems with several localized single-particle states.
We consequently focus on relative antidot diameters above the
single-level regime of the single defects ($d/\Lambda>0.42$), see
Fig.~\ref{fig:defectstates}.}
\end{thebibliography}
\end{document}